\documentclass[12pt]{article}

\title{Anomalies  in experimental data for the EPR-Bohm experiment:
Are both classical and quantum mechanics wrong?}
\author{Guillaume Adenier and Andrei Khrennikov\\
International Center for
Mathematical Modeling \\
in Physics and Cognitive Sciences\\
University of V\"axj\"o, Sweden\\
Email:Andrei.Khrennikov@msi.vxu.se}

\begin{document}
\maketitle

\abstract{We analyze anomalies in data to test the violation of
Bell's inequality for the EPR-Bohm experiment. We found that the
experimental correlations  for photon polarization have an
intriguing property. In the experimental data there are visible
non-negligible deviations of probabilities
$P_{++}^{\rm{exp}}(\alpha, \beta), P_{+-}^{\rm{exp}}(\alpha, \beta),
P_{-+}^{\rm{exp}}(\alpha, \beta), P_{--}^{\rm{exp}}(\alpha, \beta) $
from the predictions of quantum mechanics, namely, $P_{++}(\alpha,
\beta)=P_{--}(\alpha, \beta)= \frac{1}{2}\cos^2(\alpha-\beta)$ and
$P_{+-}=P_{-+}(\alpha, \beta)=\frac{1}{2}\sin^2(\alpha-\beta).$
However, in some mysterious way those deviations compensate each
other and finally the correlation $E^{\rm{exp}}(\alpha, \beta)=
P_{++}^{\rm{exp}}(\alpha, \beta)- P_{+-}^{\rm{exp}}(\alpha, \beta)-
P_{-+}^{\rm{exp}}(\alpha, \beta)+ P_{--}^{\rm{exp}}(\alpha, \beta)$
is in the complete agreement with the QM-prediction, namely,
$E(\alpha, \beta)= P_{++}(\alpha, \beta)- P_{+-}(\alpha, \beta)-
P_{-+}(\alpha, \beta)+ P_{--}(\alpha, \beta)= \cos 2(\alpha-\beta).$
Therefore such anomalies play no role in the Bell's inequality
framework. Nevertheless, other linear combinations of experimental
probabilities do not have such a compensation property. There can be
found non-negligible deviations from predictions of quantum
mechanics. Thus neither classical nor quantum model can pass the
whole family of statistical tests given by all possible linear
combinations of the EPR-Bohm probabilities. Does it mean that both
models are wrong?}

\section{Introduction}
In this  note we  continue the discussion \cite{GA}, \cite{GA1} on
anomalies in statistical data obtained in the experimental test
\cite{Weihs} that showed the violation of Bell´s inequality
\cite{bell64} and closed the locality loophole. These anomalies were
discovered in \cite{GA}, \cite{GA1}, cf. also with anomalies
discussed in the PhD-thesis of Alain Aspect \cite{Asp}. We found
that, although the experimental data really confirm the
QM-prediction for correlations:
\begin{equation}
\label{B1} E(\alpha, \beta)= P_{++}(\alpha, \beta)- P_{+-}(\alpha,
\beta)- P_{-+}(\alpha, \beta)+ P_{--}(\alpha, \beta),
\end{equation}
the QM-predictions can be violated for other linear combinations of
experimental probabilities which are different from $E(\alpha,
\beta).$

This discovery of mentioned anomalies  has extremely important
consequences for the whole Bell's program  of confronting classical
and quantum models through the statistical test based on
correlations. New advanced experiments should be performed to make a
conclusion on applicability of Bell's scheme.

\section{Anomalies: What is special in correlations?}

We have seen \cite{GA}, \cite{GA1} that the correlation is a very
special linear combination of probabilities which is surprisingly
stable with respect to deviations of its summands from the
QM-predictions. Although the experimental data can show deviations
from QM for summands in the expression (\ref{B1}), these deviations
{\it compensate each other} in the linear combination (\ref{B1}).
Finally, (in spite of mentioned deviations for terms) the
experimental correlation
$$E^{\rm{exp}}(\alpha, \beta)= P_{++}^{\rm{exp}}(\alpha, \beta)-
P_{+-}^{\rm{exp}}(\alpha, \beta)- P_{-+}^{\rm{exp}}(\alpha, \beta)+
P_{--}^{\rm{exp}}(\alpha, \beta)$$is in the agreement with
predictions of QM. Hence, for some angles, the Bell's inequality is
violated and the classical probabilistic model which was proposed by
J. Bell to confront QM should be rejected.

In contrast to many papers, see, e.g., a number of papers in
\cite{V1}, we do not worry about rejection of the classical model
for the EPR-Bohm experiment that was used by J. Bell. We {\it agree
that it should be rejected.}\footnote{ We emphasize that there can
be proposed various classical probabilistic  models for the EPR-Bohm
experiment which are different from the Bell's one. For some of such
classical models, Bell's inequality does not hold, see, e.g.,
\cite{V1}, \cite{V2}. But we do not discuss this problem in this
paper.} We worry that, as it was pointed out in introduction, other
linear combinations of probabilities do not exhibit such a
deviation-compensation property.

One might think that any linear combination $$E_c(\alpha, \beta)=$$
\begin{equation}
\label{B2} c_{++} P_{++}(\alpha, \beta)+ c_{+-} P_{+-}(\alpha,
\beta)+ c_{-+}P_{-+}(\alpha, \beta)+ c_{--}P_{--}(\alpha, \beta),
\end{equation}
where $c=(c_{++},c_{+-}, c_{-+},c_{--})$ is a real vector, has the
same deviation-compensation property. However, we found in
\cite{GA}, \cite{GA1} that it was not the case. Depending on the
choice of the vector of coefficients $c$, the linear combination
$$E_c^{\rm{exp}}(\alpha, \beta)=$$ $$
c_{++}P_{++}^{\rm{exp}}(\alpha, \beta)+
c_{+-}P_{+-}^{\rm{exp}}(\alpha, \beta)+
c_{-+}P_{-+}^{\rm{exp}}(\alpha, \beta)+
c_{--}P_{--}^{\rm{exp}}(\alpha, \beta)$$ can violate predictions of
quantum mechanics (because there is no more compensation of
deviations exhibited by individual terms).

We do not know the answer to the question in the title of this
section. One could not reject the possibility that there can be
found purely statistical reasons for the surprising stability of
$E^{\rm{exp}}(\alpha, \beta)$ to deviations in individual EPR-Bohm
probabilities.

We neither exclude the possibility that a source of compensations is
in the experimental arrangement of the EPR-Bohm tests. Thus a
detailed analysis of the experimental arrangement is required.

One of the most natural explanations is that in the real experiment
one does not prepare a singlet state, but something else. We tried
to explore such an explanation in \cite{GA}, \cite{GA1}. We started
with the hypothesis that the experimental state is pure, but {\it
not maximally entangled.} However, we found some linear combinations
$E_c^{\rm{exp}}(\alpha, \beta)$ which deviate even from the
QM-predictions for nonmaximally entangled states.

\section{Bell's arguments would imply that both classical and
quantum models should be rejected}

It is well known that the experimental statistical data, see, e.g.,
\cite{Asp} and \cite{Weihs}, violates predictions of the classical
model proposed by J. Bell to confront quantum mechanics, namely, the
Bell's inequality is violated. On the basis of the statistical test
given by the Bell's inequality this classical model should be
rejected. We completely agree with this result of the experimental
research.

However, we found that the same data violates the QM-predictions for
some tests $E_c(\alpha, \beta).$ Should one also conclude that the
quantum model should be rejected?

If we follow Bell's reasoning then the quantum model should be also
 rejected:

\medskip

There is a family of statistical tests $E_c(\alpha, \beta).$ We have
the experimental data.  Any model which does not pass one of these
tests should be rejected.  Thus we have no other choice than to
reject any model which does not pass another test.

\medskip

{\bf Conclusion:} {\it If we follow Bell's reasoning then both
classical and quantum models should be rejected on the basis of the
present experimental statistical data.}

\section{New experiments}

The crucial question is:

\medskip

{\it Can one perform an experiment that will produce data confirming
predictions of QM for all statistical tests $E_c(\alpha, \beta)$ at
the same time?}

\medskip

If one succeed in performing such a ``super-experiment'', then the
Bell's approach to confronting classical and  quantum models would
be justified.

If it is impossible to perform such a ``super-experiment'', then we
should seriously question the whole Bell's approach.

In any event one of the definite consequences of our analysis is
that the complete experimental data should be available for
theoreticians.

\end{document}